# Model of Controlled Synthesis of Uniform Colloid Particles: Cadmium Sulfide

*Sergiy Libert, Vyacheslav Gorshkov, Dan Goia, Egon Matijević and Vladimir Privman\**

*Center for Advanced Materials Processing,*
*Departments of Chemistry and Physics,*
*Clarkson University, Potsdam, New York 13699, USA*

\*Corresponding author, e-mail: privman@clarkson.edu

**Running Title:** Model of Controlled Synthesis of Uniform Colloidal Particles

**ABSTRACT**: The recently developed two-stage growth model of synthesis of monodispersed polycrystalline colloidal particles is utilized and improved to explain growth of uniform cadmium sulfide spheres. The model accounts for the coupled processes of nucleation, which yields nanocrystalline precursors, and aggregation of these subunits to form the final particles. The key parameters have been identified that control the size selection and uniformity of the CdS spheres, as well as the dynamics of the process. This approach can be used to generally describe the formation of monodispersed colloids by precipitation from homogeneous solutions.

**KEYWORDS:** Cadmium sulfide monodispersed; Colloid formation; Nanoparticle aggregation; Rate equations; Nucleation

# 1. Introduction

In our earlier studies of synthesis of monodispersed cadmium sulfide [1] and gold [2-4] spherical colloid particles, a two-stage precipitation model [1-3,5] was utilized to describe the size selection mechanism. Formation of colloids of controlled narrow size distribution is important in many applications, and its mechanism has received much attention [1,6-20], resulting in various models that yield peaked particle size distribution [9-23]. It has been experimentally documented [8,24-33] that the precipitated uniform particles are typically polycrystalline, and that their formation involves two distinct dynamical stages. In the first process, nanosize crystalline precursors (primary particles) are nucleated in supersaturated solution, while in the second process, these primary particles aggregate into larger colloids (secondary particles).

Earlier versions [1-3] of the two-stage model, when applied to gold and cadmium sulfide systems, were successful in predicting semi-quantitatively the position of the peak-shaped experimental size distributions, and, at early times, their widths. In the present work, the model is refined, which leads to better agreement between experimental data and theory, and further clarifies the physical mechanisms of particle formation. Due to the availability of detailed experimental results reported in the companion article [34], it was possible to test the validity of the model for different experimental regimes.

Originally [2,3], the effective surface tension, $\sigma$ of the nanosize primary particles was identified as a parameter that significantly influences the process. However, the average secondary particle size predicted was smaller than the experimental values. In a more recent formulation, the origin of this discrepancy was traced to the generation of too many secondary particles, and a suppression of the dimer formation [1] was introduced to account for the particle sizes, but has yielded distributions that were too narrow as compared to the experimental data.

In the present work, particle-particle aggregation of secondary particles up to a cutoff size was included, which is a new adjustable parameter, replacing the earlier dimer-suppression factor. Allowing for cluster-cluster aggregation, in addition to the cluster-singlet aggregation [1], results in broader size distributions of the final colloidal particles, consistent with the experimental findings. This approach is discussed and substantiated in Section 2, where the improved model is introduced. Numerical procedure and results of fitting the experimental data are presented in Section 3.

## 2. Model description

The model of formation of uniform polycrystalline colloid particles [1-3,5] incorporates two dynamical processes. The first process is nucleation of primary particles in supersaturated solution. The primary particles are nanosize crystals, formed by the burst-nucleation process [35,36] and further growth by diffusive capture of solute species. The second process modeled, is the aggregation of these precursors to form the secondary particles of colloidal dimensions. The rate of the primary-particle formation per unit time and volume, $\rho(t)$, is the quantity needed to connect the two processes.

The rate $\rho(t)$ depends on the degree of supersaturation in the solution, $\eta = c(t)/c_0$, where $c_0$ is the saturation concentration, $c(t)$ is the concentration of molecules (or atoms) of the precipitating material at a given time. The function $c(t)$ can be found by solving a nonlinear differential equation,

$$dc/dt = -\psi(c) + \xi(t) \ . \tag{1}$$

Here $\psi(c)$ represents the decrease in the concentration of the precipitating solute (here CdS), due to the formation of primary particles, while $\xi(t)$ is the rate of this CdS solute-species formation (see below), which depends on the conditions of the experiment.

As described in [34], CdS particles were produced in homogeneous solutions according to the following scheme:

$$C_2H_5NS \longrightarrow S^{2-} + \text{byproducts} \ , \tag{2}$$

$$Cd^{2+} + S^{2-} \longrightarrow CdS \quad \text{(solute)} \ , \tag{3}$$

$$n\,CdS \longrightarrow (CdS)_n \quad \text{(solid)} \ , \tag{4}$$

where step (2), the decomposition of $C_2H_5NS$ (thioacetamide, TAA), involves elevated temperatures and acidity, the latter controlled by adding nitric acid. This process is rate-controlling for the availability of CdS as a solute species; step (3) is assumed to be instantaneous. The release of sulfide ions by decomposition of TAA thus controls the rate $\xi(t)$ of CdS species supply.

Experimental determination of the concentration of TAA, $W(t)$, was carried out [34] by spectrophotometric analysis: the difference

$$S(t) = W(0) - W(t), \tag{5}$$

gives the concentration of $S^{2-}$, and CdS, via step (3), which proceeds as long as $Cd^{2+}$ is not exhausted. The standard energy-barrier Boltzmann-factor rate equation was assumed for the decomposition reaction in step (2),

$$dW/dt = -\Gamma e^{-\Delta E/kT(t)} W, \tag{6}$$

where $T(t)$ is the temperature profile in the experiment, which is measured directly [34]. Note that the assumption of fast step (3) corresponds to $\xi(t) = dS/dt = -dW/dt$, in Eq. (1).

The rate of consumption of the solute CdS species in the nucleation process of nanosize crystalline primary particles, $\psi(c)$, in Eq. (1), depends on the supersaturation level, $\eta(t)$, of the solution, and is given [1-3] by the relation

$$\psi(c) = \lambda c^2 (\ln \eta)^{-4} \exp[-\gamma (\ln \eta)^{-2}], \tag{7}$$

$$\lambda = 6.19 \cdot 10^4 a^9 \sigma^4 (kT)^{-4} d, \tag{8}$$

$$\gamma = 294 a^6 \sigma^3 (kT)^{-3} d, \tag{9}$$

where $a = 1.3 \cdot 10^{-10}$ m is the effective radius (corresponding to the unit-cell volume) of the solute species [37], $T$ is the temperature of the system, $d$ is the diffusion constant of the solute species, estimated from the Stokes-Einstein relation, and $\sigma$ is the effective surface tension, which is expected to be close to the bulk value. Since the value of $\sigma$ is not known for CdS, it was estimated [1] as an adjustable parameter, with the result $\sigma \approx 0.48$ N/m, a reasonable order of magnitude for minerals [38].

Of the six data sets obtained under various conditions in the experimental work [34], the size distribution as a function of time was measured for three systems (schemes T4, T5, T6 in [34]) that yielded larger particles, amenable to the size analysis by scanning electron microscopy. In each case about 150 particles were counted. These three data sets have differed by the initial concentration of TAA, $W(0)$, which resulted in different rates of supply of the solute CdS.

The time dependence of the temperature was also controlled in the experiment [34], by increasing it from 25°C up to 92°C during the first 11 min, and then keeping it approximately constant [34]. The decomposition of TAA for all three cases was modeled as follows. For the initial times, up to $t = 7$ min, the supply of the sulfide ions was negligibly small, and the TAA concentration was assumed constant. As the temperature increased, from 7 min to the final time of 20 min, Eq. (6) was used with the experimentally measured $T(t)$. A reasonable data fit was obtained with the parameter values $\Gamma \approx 3.77 \cdot 10^{15} \sec^{-1}$ and $\Delta E / k \approx 1.54 \cdot 10^4$ K, as shown in Figure 1. These values, i.e., large $\Gamma$ and $\Delta E / kT(t) \gg 1$, actually suggest that the origin of the temperature dependence of the rate of TAA decomposition may not be due to activation over an energy barrier; however, this issue is outside the scope of the present work.

The rate of the primary particles production, $\rho(t)$, can be evaluated [1-3] from the relation

$$\rho(t) = \psi(c) \left( \frac{3kT \ln \eta}{8\pi a^2 \sigma} \right)^3, \tag{10}$$

where the expression multiplying $\psi(c)$ is the number of molecules in the critical nucleus. The primary particles can further grow and aggregate into secondary particles.

The growth of the secondary particles by aggregation, is described by the relations

$$\frac{dN_{s>1}}{dt} = \frac{1}{2} \sum_{m=1}^{\min(s_{max},s-1)} \gamma_{m,s-m} N_m N_{s-m} - N_s \sum_{m=1}^{M} \gamma_{s,m} N_m \; , \qquad (11)$$

where $M = \infty$ for $s_{max} \geq k > 1$, and $M = s_{max}$ for $k > s_{max} \geq 1$,

$$\gamma_{m,n} = \alpha (m^{1/3} + n^{1/3})(m^{-1/3} + n^{-1/3}) , \qquad (12)$$

$$\frac{dN_1}{dt} = \rho(t) - 2\gamma_{1,1} N_1^2(t) - N_1(t) \sum_{k=2}^{\infty} \gamma_{1,k} N_k(t) . \qquad (13)$$

Here $N_s(t)$ is the concentration of the secondary particles consisting of $s$ primary particles, $\alpha = 4\pi r D \zeta$ is the diffusional-capture-of-particles Smoluchowski rate coefficient [39], with $r$ and $D$ denoting the radius and diffusion constant of the primary particles, and $\zeta \approx (0.58)^{-1/3} \approx 1.2$ is the factor accounting for the void volume, where 0.58 is the typical filling fraction of the random loose packing of spheres [1,2]. These equations allow for pair-wise aggregation of particles, at least one of which contains less than the threshold number of primary particles, $s_{max}$, to be discussed shortly.

The secondary particles are counted according to the number of the crystalline subunits in them. These subunits can be of varying size. In fact, the size distribution of the primary particles is rather wide [2]. The details of their growth are not accounted for in the model [1-3], because the rate expression, involving the product of the radius and diffusion constant, is not sensitive to the precise size of diffusing particles. Instead, the experimentally measured subunit size was used as the value of *r*, for each data set [34].

The process of aggregation of two particles involves attachment, merging and surface restructuring. Experimentally, it has been observed that only the smallest secondary particles are active as diffusing building blocks in the aggregation process. In fact, earlier studies have largely assumed [1-5,9-11] that only singlet (primary) particles can attach to secondary particles (and to each other). This assumption, however, results in size distributions that are narrower than those experimentally measured. Since cluster-cluster aggregation is generally known to broaden the distribution, the maximal secondary-particle size for attachment, determined by the threshold number of subunits, $s_{max}$, has been used here as a new adjustable parameter. The origin of the cut-off $s_{max}$ is dynamical, and it is likely not sharp, but we use this approximation to have only a single new adjustable parameter.

For computer simulations, the rate equations were converted to continuum differential equations, as described in [1], and solved on a time-size grid. As indicated by the experimental data, the number of the primary particles in a typical secondary particle can reach about $2 \cdot 10^6$. The employed numerical procedure has evaluated the secondary particle size distribution with the *s*-grid of up to $3 \cdot 10^4$ nodes. The *s*-step was increased with the *s* value, in a geometric progression, with ratio 1.00024. With time steps of order $10^{-3}$ sec, the total CPU time for one parameter set did not exceed 7 hours on a 1.6 GHz PC.

## 3. Results and discussion

The results of the numerical simulations, along with the experimental data for different setups described in [34], are presented in Figures 2-4. The best possible fit for all the systems considered, was achieved with $s_{max} = 25$ and $\sigma = 0.55$ N/m, when these two quantities were varied as the adjustable parameters, both being of reasonable magnitude. The value for the surface tension, $\sigma = 0.55$ N/m, is higher than estimated in an earlier study [1], $\sigma = 0.48$ N/m, using a somewhat different model, and it is still within a reasonable range for minerals. As mentioned, the present model has the advantage of reproducing quite reasonably the width of the distribution (Figs. 2-4), by allowing cluster-cluster aggregation, rather than suppressing very small particle aggregation, as tried earlier [1,3]. The variation of the estimated surface tension suggests that the error bar for $\sigma$ is approximately 15 to 20 %.

The value of $s_{max} = 25$ is also reasonable, corresponding to at most 2 shells in a random packing of nanosize crystalline primary particles as subunits in those secondary particles that still are mobile enough and small enough to be able to diffuse to other secondary particles and restructure to merge into one object. As expected, when the value of $s_{max}$ was increased, allowing for cluster-cluster aggregation of larger secondary particles, the resulting peaks became broader and shifted to larger sizes, at given times.

As in earlier model studies [1-3], the results are particularly sensitive to the value of the surface tension $\sigma$, which can be summarized by the following observations. First, the most important from the point of view of the dynamics of the process is the effect of the surface tension on the relative spacing between peak positions at different times. If the value of $\sigma$ is decreased, the rate of the peak-position growth decreases. Increasing the value of $\sigma$ generally increases the rate of the peak-size growth. Note that this rate is not constant, and the peak ultimately freezes at a certain size value, so the model captures the "size selection" in colloid synthesis. Secondly, the value of $\sigma$ significantly affects the peak formation and its width. Generally, increasing $\sigma$ causes the initial width of the peaked distribution to broaden, with the results that a smaller number of larger particles is produced.

The evolution of the secondary particles size distribution, $N_s(t)$, is graphically represented in Figure 5. The supply of the new solute species is the primary external control mechanism; after all, one cannot control $\sigma$, though $s_{\max}$ can be influenced by the chemical composition of the solution and the nature of the stabilizer. The rate at which the solute species are available determines the rate of production of the primary particles, namely, clusters with $s = 1$. One can consider two groups of the growing clusters (secondary particles): those that are smaller than $s_{\max}$, and those that exceeded the threshold $s_{\max}$. The latter particles cannot attach to each other and the sizes of most of them are in the peak. The smaller ones can either directly attach to the larger particles, within the growing peak or its shoulder, or they can grow more slowly, between $s = 1$ and $s_{\max}$, by aggregating with other small particles. In the early stages, the growth process proceeds via the latter mechanism, that of small-particle aggregation. The numerical simulations carried out indicate that, at a certain time, the former pathway, that of capture by the particles in the peak, becomes dominant. Once this regime is reached, most of the newly supplied particles will be consumed by the peak, which thus grows without changing the number of particles in it.

If this peak-evolution stage of the process, described in [5], takes over when the number of particles larger than $s_{\max}$ is small, then they will grow to larger sizes. Thus, for the system T4 [34], the initial concentration of TAA was low, causing a lower rate of sulfide ions release, and consequently lower rate of primary particles supply. This caused the system to switch into the peak-growth regime with smaller number of particles that exceed $s_{max}$, resulting in the formation of larger final spheres (Figure 2), as compared to T5 and T6 (Figures 3 and 4). For the initially higher rates of the primary particles supply, the peak-growth regime is achieved for larger number of particles over $s_{max}$, yielding smaller final colloids (scheme T6 in [34], Figure 4).

Let us consider now the effects of the system parameters and synthesis schemes on the width of the final particle size distribution. The influence of the value of $\sigma$ on the width of the peaked size distribution can be summarized as follows. Increasing the $\sigma$ value causes the rate of the primary particles formation to decline, because increased surface energy suppresses nucleation. This lower rate of the primary particle formation actually causes the width of the distribution to increase. As observed earlier, the system should switch into the peak-growth regime, when most of the supplied particles are attaching directly to the particles in the peak. This occurs when the total surface area of all the particles larger than $s_{max}$ exceeds some critical value. At a lower initial rate of subunit supply, this would be achieved at later times, when the size distribution of particles larger than $s_{max}$ has developed a significant shoulder due to the other growth mechanism, that of small particles reaching the threshold $s_{max}$ by aggregation with other small particles. Thus, the formation of the growing peak, when this becomes the dominant process, starts from a broader distribution, and consequently the final width of the peaked size distribution is relatively large.

Similarly, for systems with a low rate of solute-species supply (T4, Figure 2), the width of the growing peak is larger. In contrast, when the rate of the primary particle supply is high, either due to high concentration of TAA (scheme T6, Figure 4) or low value of $\sigma$ (checked by numerical simulations), the peak-growth regime is achieved at earlier times, with many $s > s_{max}$ particles in the system, causing narrower size distribution, yet smaller particle size, for the same total amount of the solute species supplied.

In order to produce highly monodisperse colloidal particles, one would want to have as few particles as possible in the shoulder: the region from $s_{max}$ to approximately the beginning of the peak. If this condition is not achieved, a significant fraction of the supplied particles would be consumed by the shoulder, thus broadening the peak, resulting in a wide size distribution of the final colloids. It transpires that in order to produce highly uniform colloids, one should have a relatively sharp peak already in the early stages of the process [5]. Afterwards, the matter supply rate should be kept moderate: too fast a supply of primary particles would populate the unwanted shoulder region (Figure 5).

In summary, the analysis of the experimental data for CdS colloid particle synthesis suggests that a model with cluster-cluster aggregation of secondary particles containing up to approximately 25 primary particles, can provide the desired mechanism of obtaining a smaller number of larger final particles. This model mechanism yields the width of the distribution much closer to that experimentally observed, than earlier approaches.

This research has been supported by the National Science Foundation (grant DMR-0102644) and by the Donors of the Petroleum Research Fund, administered by the American Chemical Society (grant 37013-AC5, 9).

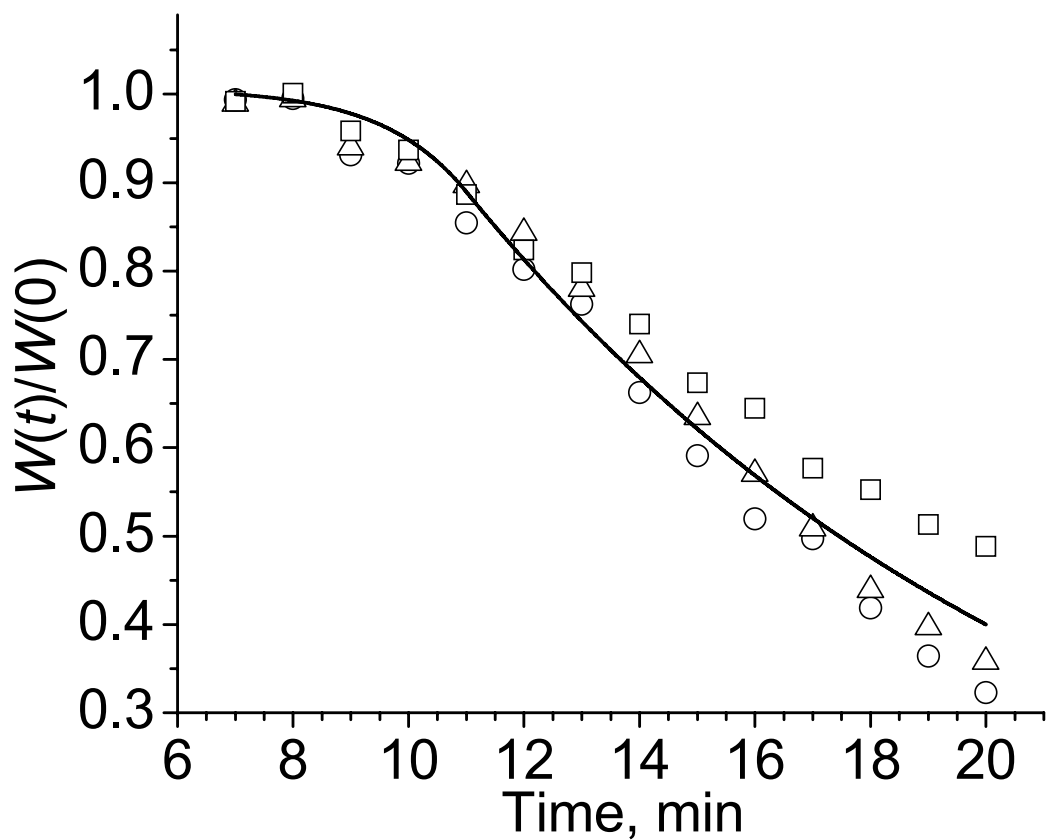

**Figure 1:** Experimental values and theoretical curve (solid line) for the TAA concentration as a function of time. The symbols □ mark the values for the system denoted T4 in [34], with the initial concentration $W(0)=0.0135$ mol·dm$^{-3}$. The symbols ○ mark the values for the system T5, with $W(0)=0.0270$ mol·dm$^{-3}$, while △ mark the values for the system T6, with $W(0)=0.0560$ mol·dm$^{-3}$.

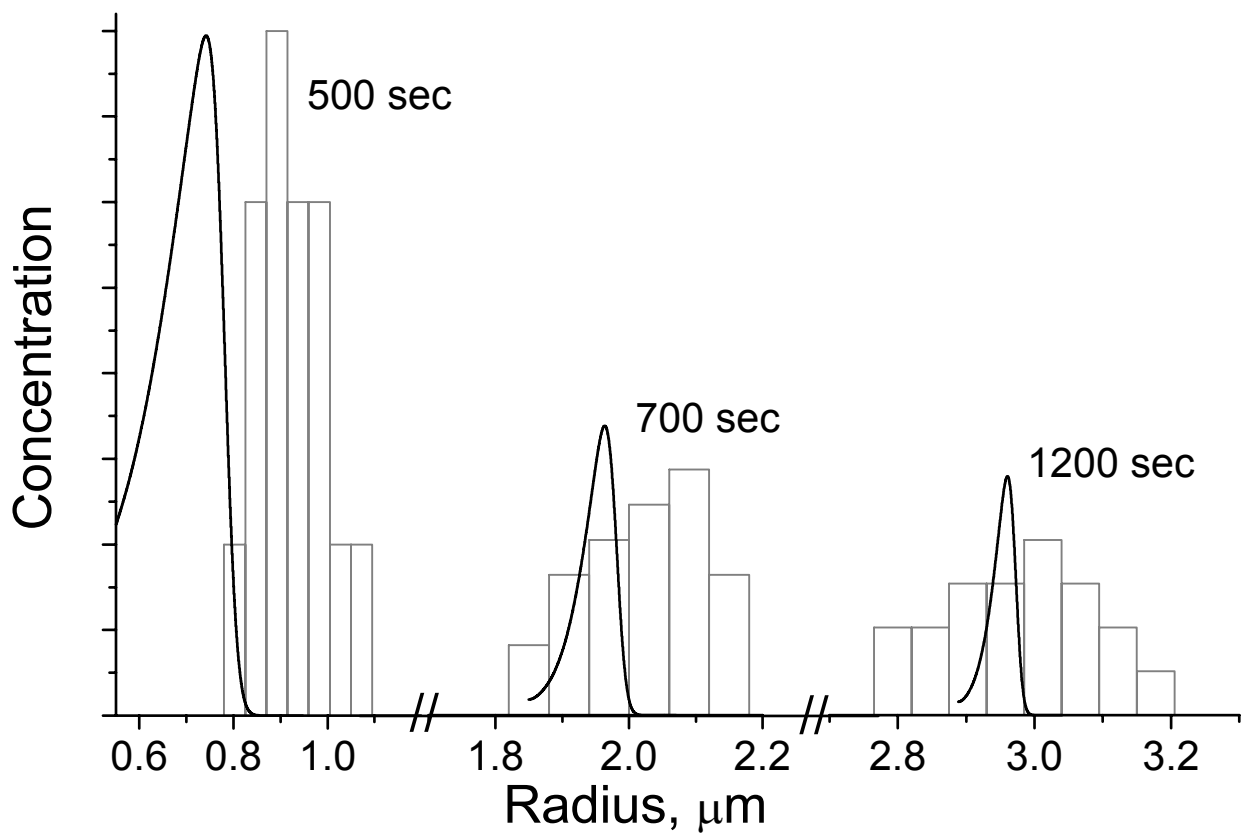

**Figure 2:** Experimentally determined histogram of size distribution of spherical CdS particles prepared according to the scheme T4, as described in [34]. The solid lines show the theoretical curves obtained by numerical solution of the model equations. The concentration is shown in arbitrary units.

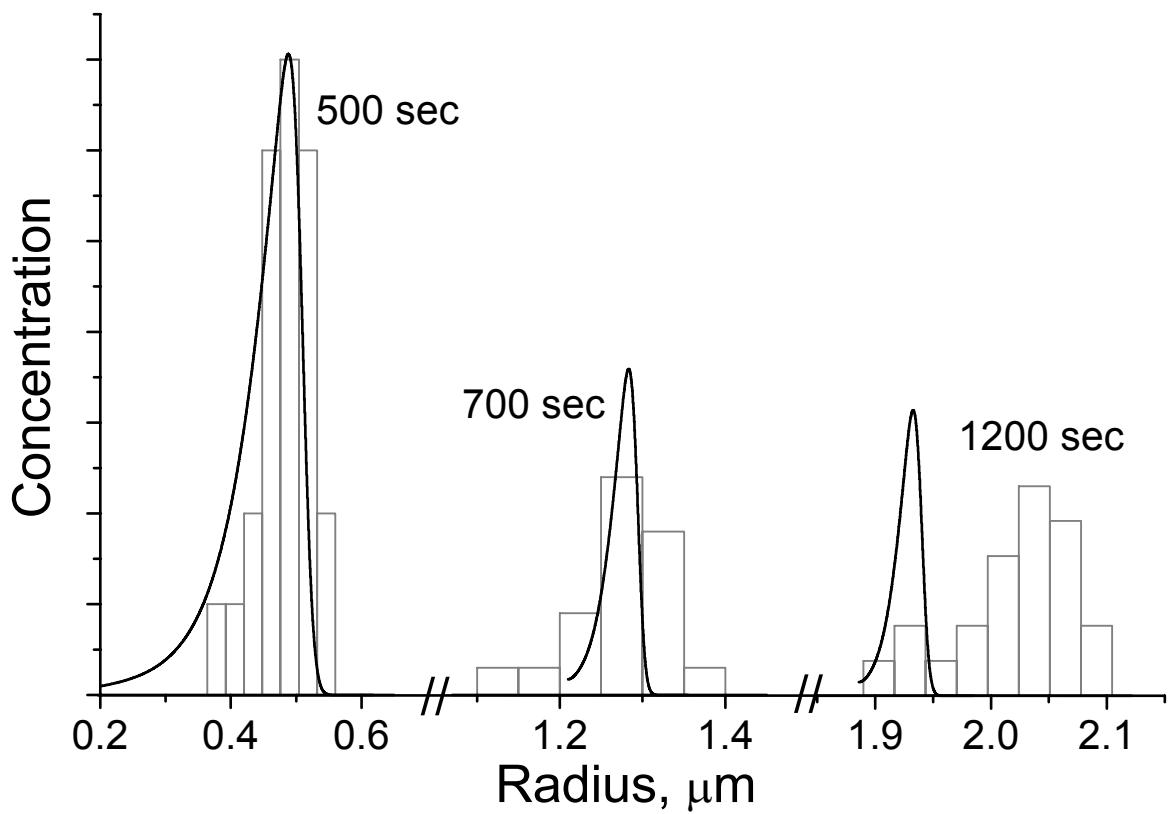

**Figure 3:** The same as in Figure 2, for the scheme T5, described in [34].

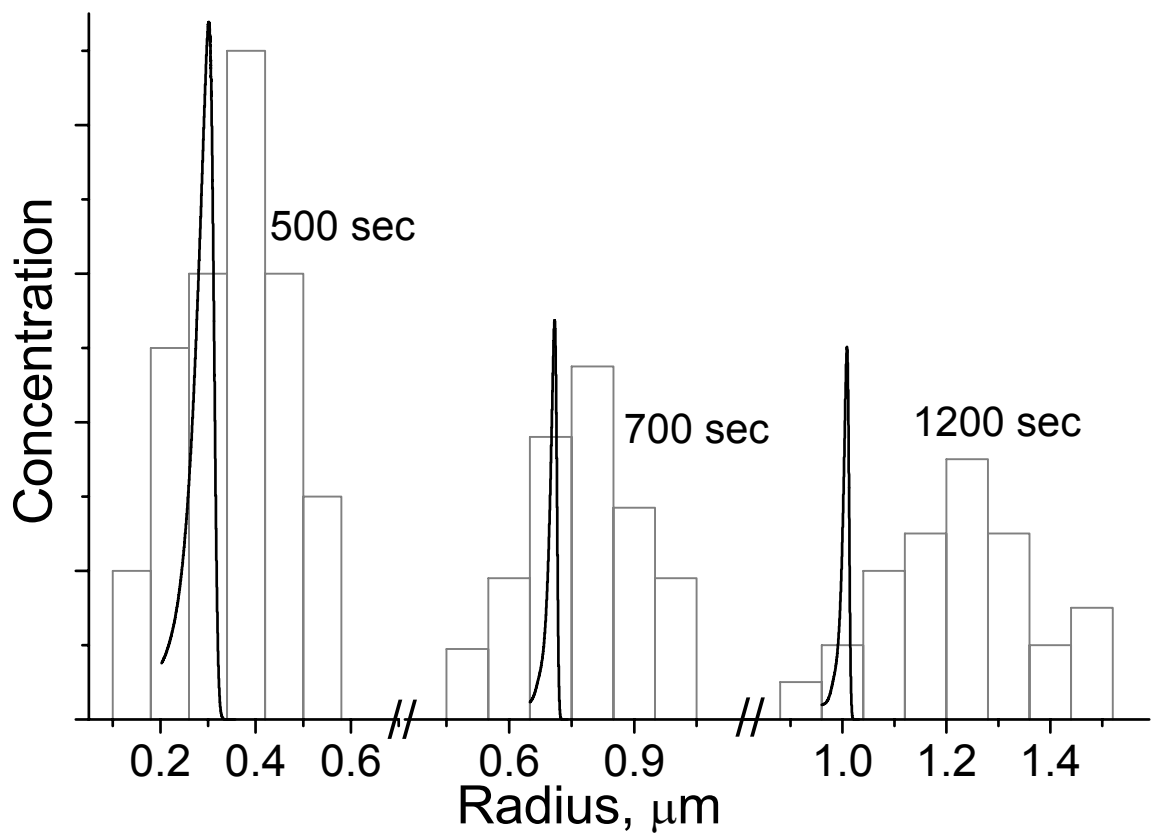

**Figure 4:** The same as in Figure 2, but for the scheme T6, described in [34].

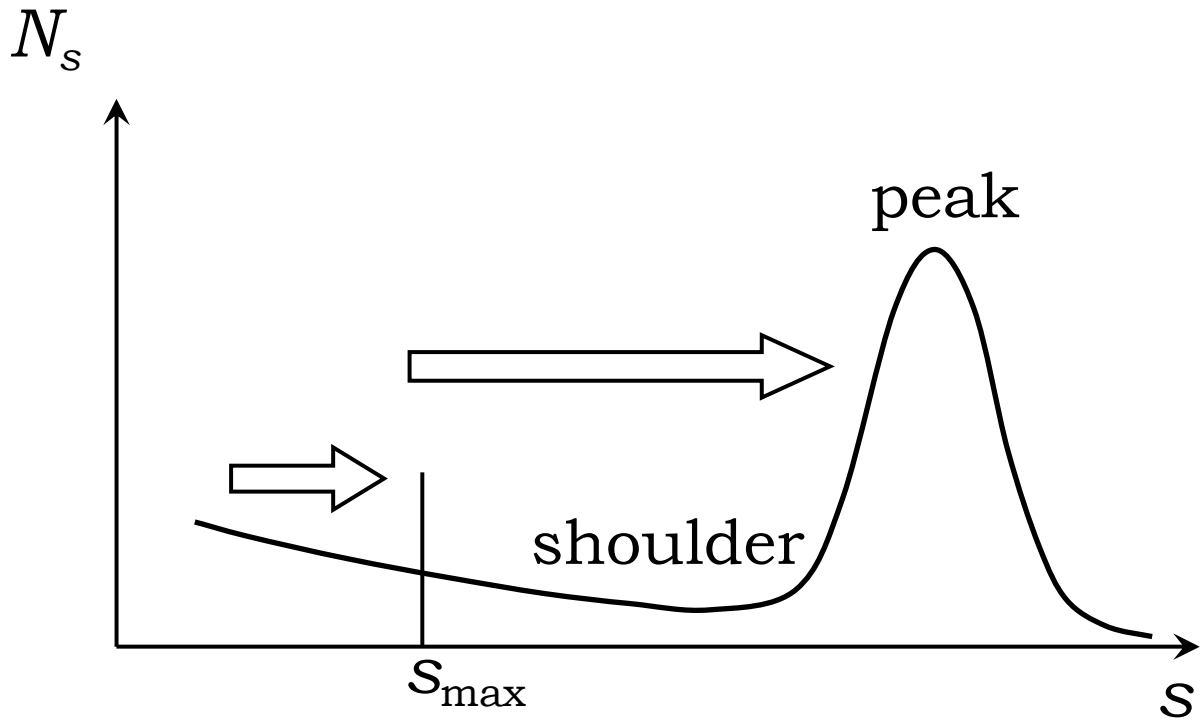

**Figure 5:** Graphical representation of the mechanism of the aggregation pathways. The system is constantly fed by new primary particles, by controlling the supply of the solute species. Two groups of secondary particles are considered. Those that are smaller than $s_{max}$, and those that have exceeded the threshold $s_{max}$ and cannot attach to each other. The smaller particles can "disappear" by direct attachment to the larger particles, primarily within the growing peak of the distribution, or they can drift along the shoulder between $s = 1$ and $s_{max}$, by aggregating with other small particles. Intensities of these two mechanisms determine the rate of the peaked size distribution formation and the degree of colloidal particle uniformity.